\documentstyle[aps,twocolumn,epsf]{revtex}

\def\be{\begin{equation}}
\def\ee{\end{equation}}
\def\bea{\begin{eqnarray}}
\def\eea{\end{eqnarray}}
\def\bml{\begin{mathletters}}
\def\eml{\end{mathletters}}
\def\bc{\begin{center}}
\def\ec{\end{center}}
\def\bfig{\begin{figure}}
\def\efig{\end{figure}}
\def\lan{\langle}
\def\ran{\rangle}
\def\si{\sigma_i}
\def\sj{\sigma_j}

\begin{document}

\title{A lattice glass model with no tendency to crystallize}
\author{M. Pica Ciamarra, M. Tarzia, A. de Candia, A. Coniglio}
\address{Dipartimento di Scienze Fisiche, Universit\`a di Napoli
``Federico II''\\
Istituto Nazionale di Fisica della Materia, Unit\`a di Napoli\\
Monte S. Angelo, via Cintia, 80126 Napoli, Italy}

\wideabs{%
\maketitle
\begin{abstract}
We study a
lattice model with two body interactions
that reproduces in three-dimensions many
features of structural glasses, like cage effect and vanishing diffusivity.
While having a crystalline state at low temperatures,
it does not crystallize when quenched,
even at the slowest cooling rate used, which makes it suitable
to study the glass transition. We study the model on the Bethe
lattice as well,
and find a scenario typical of $p$-spin models, as in the Biroli
M\'ezard model.
\end{abstract}
}

The glass transition is a phenomenon that is observed when a liquid
is cooled well below its melting temperature, avoiding the crystallization
of the system \cite{A00,DBS01}.
The motion of the molecules is slowed down more and more
by lowering the temperature, until at some point it becomes so slow
that the liquid appears as a disordered solid, with the molecules
vibrating around some equilibrium positions, but not diffusing anymore.
Two fundamental approaches have been put forward
in the study of this phenomenon, from the point of view of 
dynamics \cite{G91}, and thermodynamics \cite{MP99}.
A further insight in the understanding of the 
glass transition was achieved through
the study of the $p$-spin glass \cite{KT87}, a mean field model with
$p$-body interactions and quenched disorder, which reproduces many 
features of glass forming liquids.
More recently, Biroli and M\'ezard have
introduced a very interesting 
lattice glass model \cite{BM02}, where occupation variables
interact via $p$-body potentials.
The mean field solution on the Bethe lattice
shows a scenario identical to $p$-spin models,
and three dimension numerical simulations show a behavior typical of
glass forming liquids.
In order to reduce the tendency to crystallize, the authors
considered a mixture of two types of particles.
Further study of the model was done in \cite{L02}. Another 
model with $p$-body interactions and without quenched disorder, that
shows a similar behavior, was studied in \cite{FM01}.

In this paper we study a simple lattice gas model with 
only two body interactions. The model reproduces in mean field again
the scenario typical of $p$-spin models, and three dimension
numerical simulations show the same behavior typical of glass forming liquids,
including cage effect and two step relaxation.
Moreover, while having a crystalline phase at low temperatures,
the model does not crystallize when cooled, even in the pure 
case (only one type of particles) and
at extremely low cooling rates.
This makes the model more suitable to study the glass transition,
and allows to obtain very clear results.

The model is defined as follows. 
We partition the space in regular cells, such that not more than one particle
can have its center of mass inside the cell. 
The position and orientation of the particle inside
the cell is represented by a coarse grained discrete internal degree of
freedom, that can assume a finite number $q$ of states,
and the interaction between nearest neighbor particles depends
explicitly from their position inside the cell.
The model is therefore described by the following Hamiltonian
\be
{\cal H}=\sum_{\lan ij\ran} n_i n_j \phi_{ij}(\si,\sj)-\mu\sum_i n_i
\label{model}
\ee
where $n_i=0,1$ whether the $i$-th cell is occupied by a particle or not,
$\si=1,\ldots,q$ represents the internal position of the particle,
and $\phi_{ij}(\si,\sj)$ is the interaction energy between two
particles in cells $i$ and $j$ with internal positions $\si$ and $\sj$.

It is clear that, by choosing a sufficiently large number $q$ of internal
positions, and an opportune interaction matrix $\phi_{ij}(\si,\sj)$, one can
approximate as closely as desired any model defined in the continuum, like
for example a Lennard-Jones liquid. On the other hand, it is plausible that
a few number of internal states may be enough to catch
the fundamental characteristics of dynamics
and thermodynamics of glass forming liquids.

Here, we study a particularly simple realization
of the model (\ref{model}).
In two dimensions, we partition the space in square cells, and
subdivide each cell into four internal positions.
When a cell is occupied by a particle in a given position,
a hard-core repulsion forbids the presence of another particle
in some of the internal positions of the neighboring cells
(see Fig. \ref{lattice}).
Therefore in this case $q=4$, and the interaction $\phi_{ij}(\si,\sj)$ is zero
if the positions $\si$ and $\sj$ are ``compatible'', infinite otherwise.
The extension to three dimensions is straightforward.
In that case one partitions the space in cubic cells, and considers
six internal positions instead of four.
For every spatial dimension $d$, the model has a crystalline ground state
with density $2d/(2d+1)$, on lattices with
sides multiples of $2d+1$ lattice spacings.
For example, on a cubic lattice with periodic boundary conditions
it can be found as follows: consider the cell with coordinates $(x,y,z)$,
and evaluate the number $a=(x+2y+3z \bmod 7)$:
if $a=0$, leave the cell empty;
if $a=1,2,3$ put a particle in the negative $x$, $y$ or $z$ direction,
respectively;
if $a=4,5,6$ put a particle in the positive $z$, $y$ or $x$ direction,
respectively.

We have simulated the model in three dimensions, by means of grand-canonical
and canonical dynamics. A compression experiment
on a system of size $28^3$ is shown in Fig.\ \ref{compress}.
A simple Monte Carlo grand-canonical (variable density) dynamics
is performed, and temperature is slowly decreased,
starting from some high value. The results are shown as open circles
in Fig.\ \ref{compress}, for various cooling rates.
The final high-density state is strongly dependent from the cooling rate,
as observed in glass forming liquids. On the other hand, no tendency to
crystallization is observed: even at the slowest cooling rate
($\dot{T}/T=-10^{-7}$)
no transition to the crystalline state was observed.
We have also computed the density in the crystalline state, starting from
the ground state at very low temperature, and heating up
slowly (diamonds in Fig.\ \ref{compress}).
This ordered state becomes unstable at $T\simeq 0.150\,\mu$,
where the system falls into the liquid phase.

After having equilibrated the system at a certain density in the
liquid phase, we have
switched to a canonical (fixed density) dynamics. These simulations are
performed on lattices of size $15^3$.
The mean square displacement $\lan r^2(t)\ran$ of the particles is show in
Fig.\ \ref{relax}a. This is defined taking in account also the position
of the particle inside the cell, giving to each internal position a shift
$1/4$ of a lattice spacing with respect to the center of the cell.
In Fig.\ \ref{relax}b instead we plot the self-overlap, defined as
\be
\lan q(t)\ran={1\over N}\sum_i\lan n_i(t)n_i(0)\,
{\bbox{\si}}(t)\cdot{\bbox{\si}}(0)\ran
\ee
where ${\bbox{\si}}(t)$ are unit length vectors, pointing in one of the 
six coordinate directions, representing the position of the particle inside
the cell, and $N$ is the number of particles.
As it happens in glass forming liquids, at high density a {\em plateau}
develops both in the mean square displacement and in the self-overlap.
This is the signature of the so called ``cage effect''.
Particles are trapped in cages formed by neighboring particles,
and vibrate rapidly around some equilibrium position, giving rise to
the first decay in correlation functions. Only after a long time
they manage to escape from their cages, giving rise to the second 
final decay to equilibrium. While the time of the first step remains
finite, the time of the slow relaxation diverges when the glass transition
is approached.
In Fig.\ \ref{plaw}a it is shown the diffusivity as a function of
$\rho-\rho_c$, with $\rho_c=0.844$. It can be fitted by a power law, with
an exponent $\gamma=3.45$. In Fig.\ \ref{plaw}b it is shown the dynamical
susceptibility $\chi(t)=N(\lan q(t)^2\ran-\lan q(t)\ran^2)$, which shows
a behavior typical of $p$-spin models,
and observed also in a Lennard Jones binary mixture \cite{DFPG99},
with a peak at finite time
that grows when approaching the glass transition, signaling the presence
of dynamical heterogeneities.

We have then studied the model on the Bethe lattice,
namely a random lattice with fixed connectivity \cite{BM02,MP01,MP02}.
Each site of the lattice is subdivided in $k+1$ internal positions,
and is connected to $k+1$ randomly chosen neighbors
(see Fig.\ \ref{tree}a).
For large number of sites $N$, the lattice is locally treelike, 
but has loops of length of the order $\log N$.
Consider a ``branch'' ending on site $i$, that is a graph in which site
$i$ has only $k$ neighbors,
and let $Z_0^{(i)}$,
$Z_{\text{ext}}^{(i)}$ and $Z_{\text{int}}^{(i)}$ be the partition
functions of the branch, restricted respectively to configurations
where site $i$ is empty, occupied on the ``external''
position, and occupied on one of the $k$ ``internal''
positions (see Fig.\ \ref{tree}b). We define the ``local fields''
$a_i$ and $h_i$ acting on the site $i$ by the relations
$e^{\beta h_i}=Z_{\text{int}}^{(i)}/Z_0^{(i)}$,
$e^{\beta a_i}=(Z_{\text{ext}}^{(i)}+Z_{\text{int}}^{(i)})/Z_0^{(i)}$.
The branch can be seen as the result of the 
merging of the $k$ branches ending on the neighbor sites $j=1,\ldots,k$,
which leads to a recursion
relations for the local fields, and a free energy shift:
\bml
\label{recurr}
\bea
&&e^{\beta a_i}=e^{\beta\mu}\bigg(\prod_{j=1}^{k}{1+e^{\beta h_j}\over
1+e^{\beta a_j}}\bigg)\bigg(1+\sum_{j=1}^{k}{1\over1+e^{\beta h_j}}\bigg)\\
&&e^{\beta h_i}=e^{\beta\mu}\bigg(\prod_{j=1}^{k}{1+e^{\beta h_j}\over
1+e^{\beta a_j}}\bigg)
\bigg(\sum_{j=1}^{k}{1\over1+e^{\beta h_j}}\bigg)\\
&&e^{-\beta\Delta F}=Z_0^{(i)}/\prod\limits_{j=1}^k Z_0^{(j)}
=\prod\limits_{j=1}^k(1+e^{\beta a_j})
\eea
\eml
To evaluate the total free energy, one has to consider also
the free energy shifts when merging $k+1$ branches on a new site,
and when merging two branches, that are given respectively by
\bml
\bea
e^{-\beta\Delta F^{(1)}}&=&\prod_{j=1}^{k+1}(1+e^{\beta a_j})
+e^{\beta\mu}\sum_{p=1}^{k+1}\prod_{j\neq p}(1+e^{\beta h_j})\\
e^{-\beta\Delta F^{(2)}}&=&1+e^{\beta a_1}+e^{\beta a_2}+e^{\beta(h_1+h_2)}
\eea
\eml

The replica symmetric solution of the problem corresponds to the case
in which there is only one pure state,
and the fields do not fluctuate from site to site.
In this case, the total free energy is given by
$F=\Delta F^{(1)}-(k+1)\Delta F^{(2)}/2$,
from which one can find
the specific volume and the entropy per site, 
that are shown as solid lines in Figs. \ref{volume}
and \ref{entropy} (in the case $k=5$).
As it happens in the model studied by Biroli and M\'ezard \cite{BM02},
the entropy becomes negative for temperatures below some threshold.
A similar behavior is observed for every $k>1$.
One can also find a (replica symmetric) crystalline solution, where the
fields do not fluctuate on the single site, but are different on different
sites, and the merging of the branches is done preserving the crystalline
structure. For $k=5$, this solution appears below $T_{ms}=0.1633\,\mu$,
and becomes stable
below the melting temperature $T_m=0.1444\,\mu$.
The corresponding volume
and entropy are shown as dashed lines in Figs. \ref{volume}
and \ref{entropy}.

We have then looked for a solution at the level of 1-step replica symmetry
breaking (RSB) \cite{PMV87,M98,MP01,MP02}.
In this case, many pure states exist, and the local fields fluctuate
also on the single site. We work in the 
so called ``factorized case'', in which the probability 
distribution of the local fields
$P(a_i,h_i)$ on the single site is the same on all the sites of the lattice.
The self-consistency equation, when merging $k$ branches on a site $i$, 
now reads
\bea
P(a,h)\propto&&\int\prod_{j=1}^k da_j\,dh_j\,P(a_j,h_j)
\nonumber\\
&&\times\delta(a-a_i)\delta(h-h_i)\exp(-\beta m\Delta F)
\label{consist}
\eea
where $a_i$, $h_i$ and $\Delta F$ are functions of $a_j$ and $h_j$ via
the equations (\ref{recurr}),
$m$ is a real parameter, and the total free energy  is
\bea
F=&&-{1\over\beta m}\left\{\log\int\prod_{j=1}^{k+1}
da_j\,dh_j\,P(a_j,h_j)e^{-\beta m\Delta F^{(1)}}\right.
\nonumber\\
&&\!\!\left.\mbox{}-{k+1\over 2}\log\int\prod_{j=1}^2
da_j\,dh_j\,P(a_j,h_j)e^{-\beta m\Delta F^{(2)}}\right\}
\eea
One then has to maximize the free energy with respect to the parameter $m$,
in the interval $0\le m\le 1$.

At temperature $T=0$ the solution of (\ref{consist})
is easily found.
Indeed, in that case the fields $a_i$ and $h_i$
can only have the finite set of values
\be
\label{delta}
\begin{array}{rcl}
a_i&=&(1-k)\mu,\ldots,0,\mu\\
h_i&=&\rule{0ex}{5ex} 
\left\{\begin{array}{ll}
a_i&\text{if}\> a_i<\mu\\
\rule{0ex}{2.5ex}
0,\mu\qquad&\text{if}\> a_i=\mu\\
\end{array}\right.
\end{array}
\ee
The distribution $P(a,h)$ is then a sum of $k+2$ delta functions,
and the integral equation (\ref{consist}) gives rise to
$k+2$ algebraic equations for the weights of the delta functions.
One has to give a finite value to the parameter $\beta m\mu$, and maximize
the free energy with respect to it. For $k=5$, the maximum is
$F=-0.8265\,\mu$, for $\beta m\mu=8.078$.
At temperature $T>0$ the solution of (\ref{consist}) can be found
iteratively. We discretized the distribution $P(a,h)$ over
a domain of the plane $(a,h)$, using a fine grid with spacings
$da=dh=\mu/1024$, and starting from some initial distribution
applied iteratively equation (\ref{consist}) until the
procedure converged.
Of course one must be careful that the chosen
domain in the plane $(a,h)$ covers all the support of $P(a,h)$.
To locate the maximum of the free energy as a function of $m$, it is useful
to evaluate explicitly the derivative $dF/dm$ \cite{MP01}.
The results for the volume and entropy, and for the
parameter $m$, are shown as open circles in
Figs. \ref{volume} and \ref{entropy}, in the case $k=5$. The 1-step
RSB solution appears below $T_D\simeq 0.105\,\mu$,
and becomes stable below $T_K\simeq 0.087\,\mu$,
where $m=1$. Note that below $T_K$ the configurational entropy
vanishes, and therefore the residual entropy, at low temperatures in
the 1-step RSB solution, has to be interpreted as a ``vibrational entropy'',
due to the ``rattling'' of the particles inside their cages. This vibrational
entropy is absent in the crystalline state.

In conclusion, we have studied a lattice glass model with two
body interactions.
On the Bethe lattice, it shows the same
scenario of discontinuous spin glasses, and of other recently studied
lattice models of glass.
Moreover, it shows no tendency to crystallize when 
simulated in three dimension with a simple Monte Carlo
dynamics, which avoids the need
to introduce a mixture, as in many models of glass.

After this work was done, we learned about the work of Weigt and
Hartmann, who have studied a lattice glass model that shows a similar
behavior \cite{WH02}.

Acknowledgments --
This work was partially supported by the European TMR Network-Fractals
(Con\-tract FMR\-XCT\-980\-183), MURST-PRIN-2000, INFM-PRA (HOP), and
MIUR-FIRB-2002.

\bfig
\bc
\mbox{\epsfysize=4cm\epsfbox{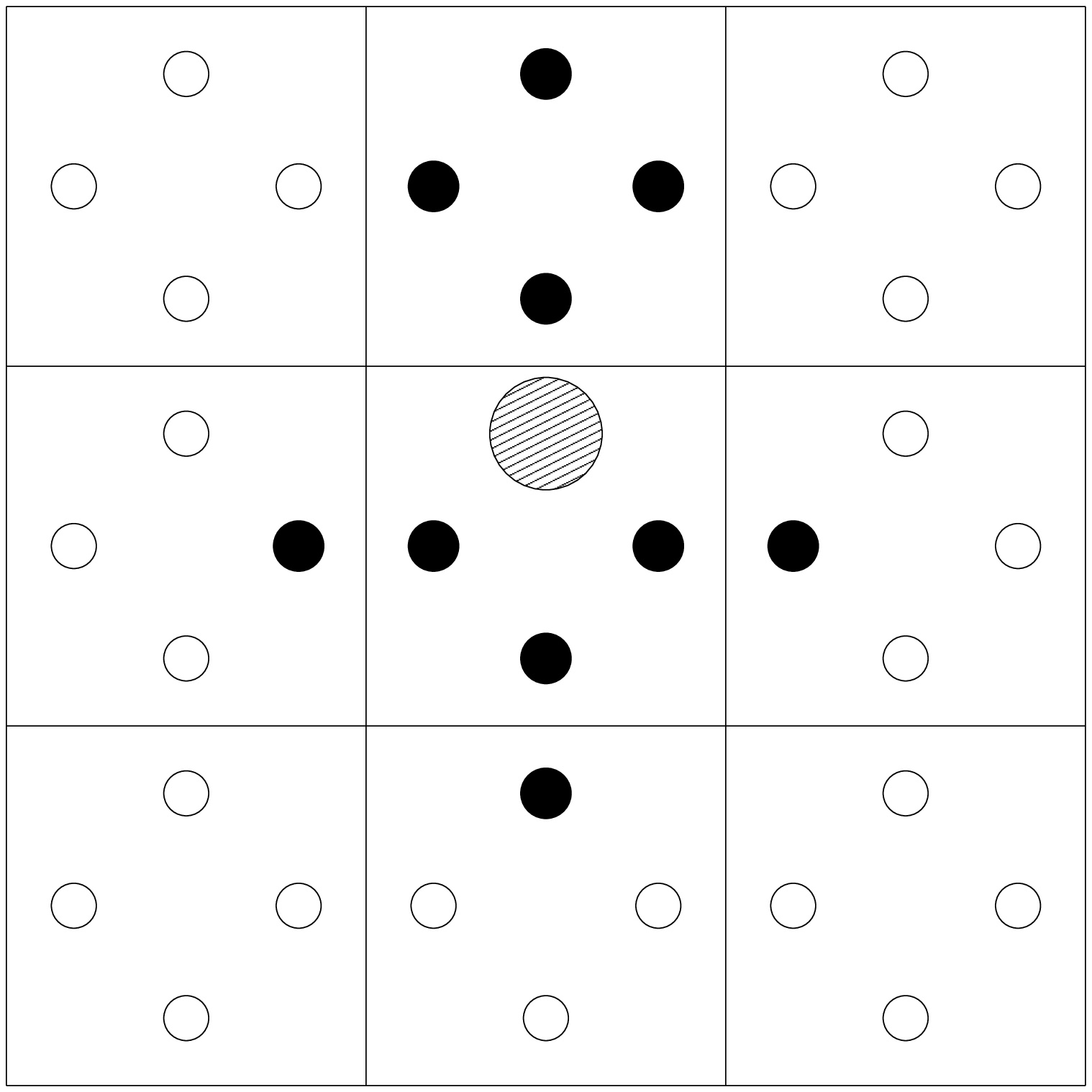}}
\ec
\caption{The model in two dimensions:
the space is partitioned in square cells, and each cell
can be occupied by at most one particle in one of four
positions (little circles). A particle in a given position
(big shaded circle) forbids the presence of another particle in the
positions colored in black.}
\label{lattice}
\efig

\bfig
\bc
\mbox{\epsfysize=5.5cm\epsfbox{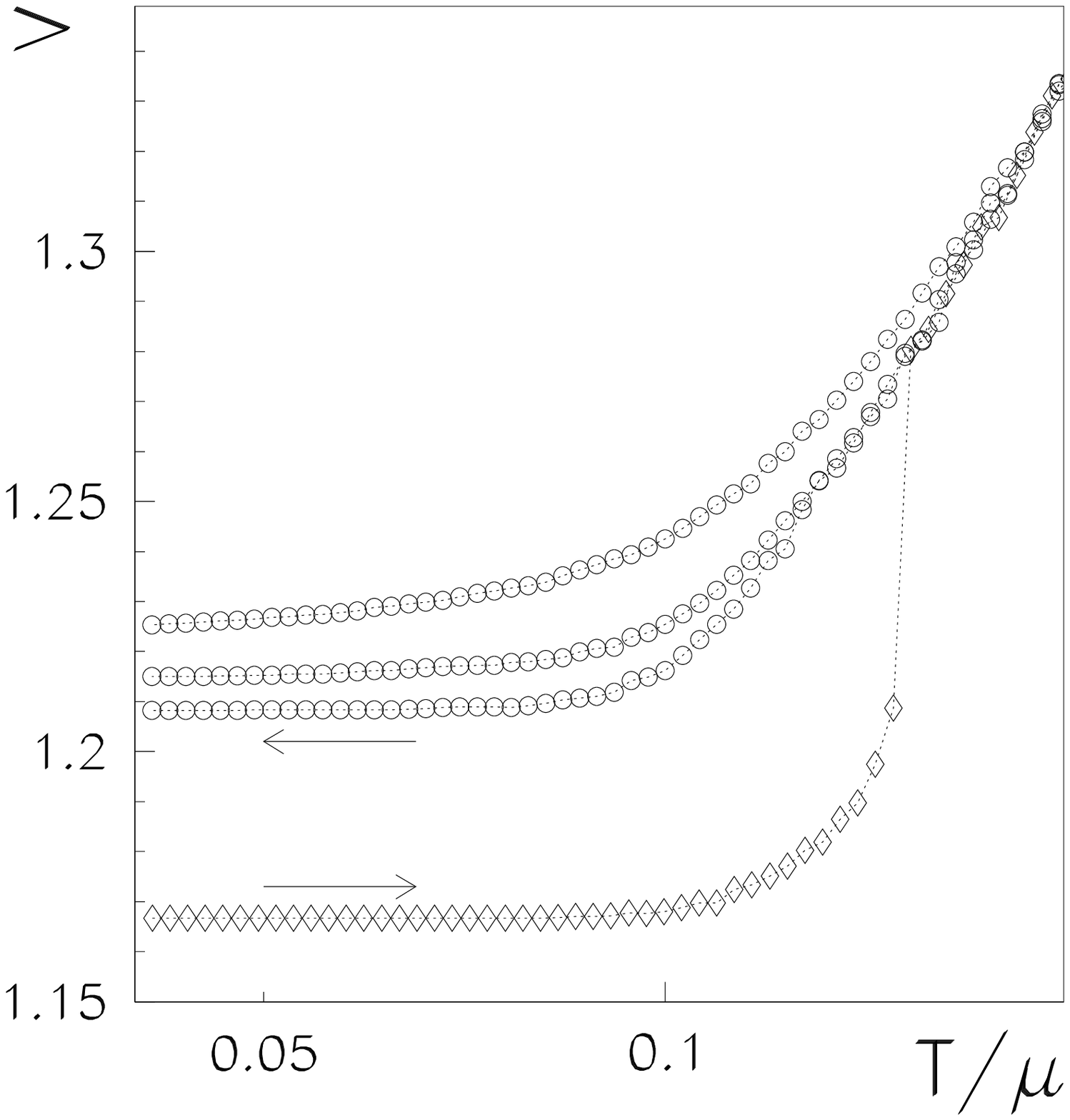}}
\ec
\caption{Specific volume (inverse density)
as a function of the temperature, for a three-dimensional
system of size $28^3$. Circles: the system is cooled
starting from high temperature, for cooling rates
(from top to bottom) $\dot{T}/T=-10^{-4}$,
$-10^{-5}$, $-10^{-7}$.
Diamonds: the system is heated starting from the crystalline ground state,
with heating rate $\dot{T}/T=10^{-7}$.
}
\label{compress}
\efig

\bfig
\bc
\mbox{\unitlength=1cm
\begin{picture}(8,4)(0.3,0)
\put(0.4,0.2){\bf a}
\put(0,0){\epsfysize=4.5cm\epsfbox{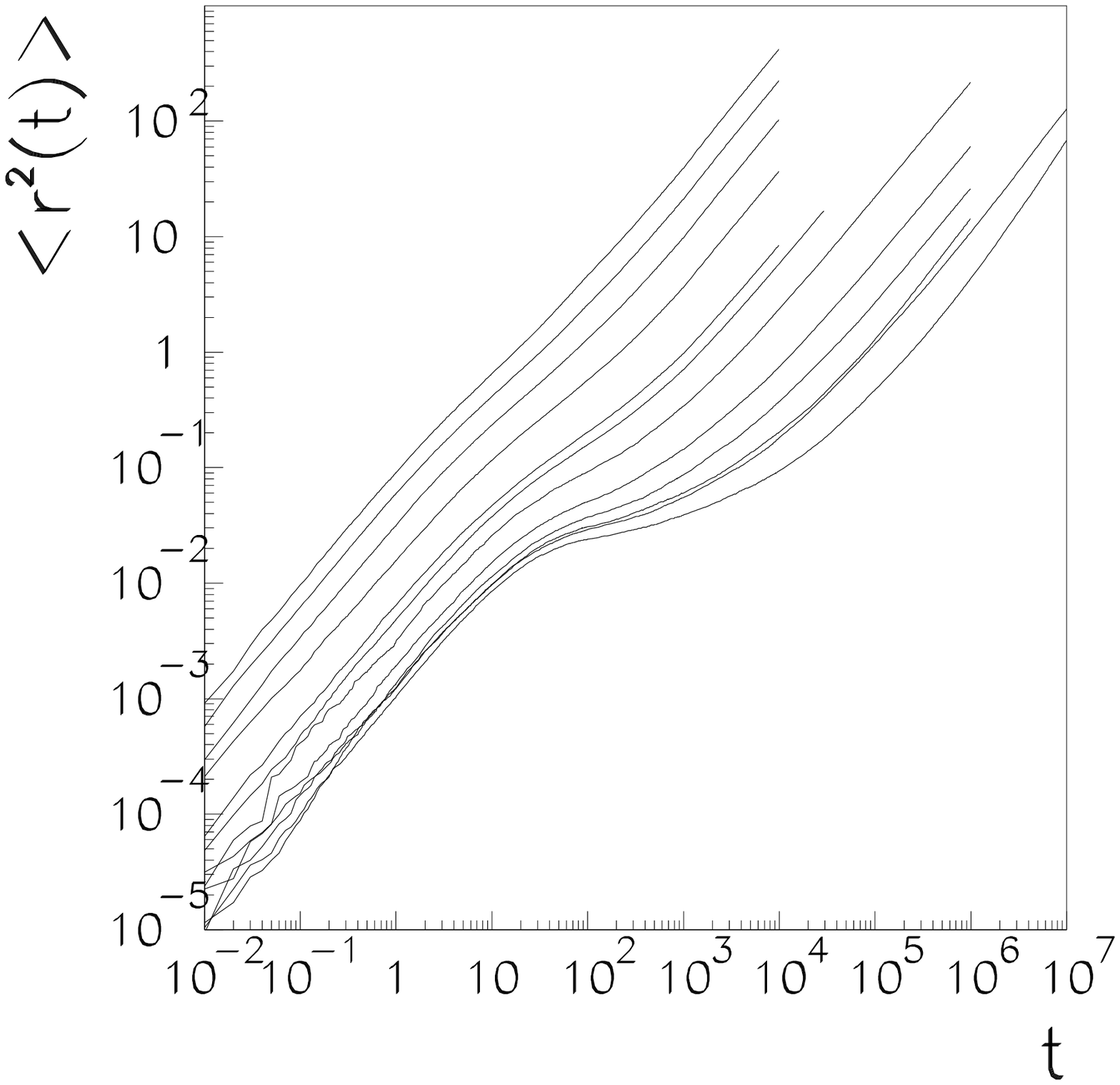}}
\put(4.8,0.2){\bf b}
\put(4.4,0){\epsfysize=4.5cm\epsfbox{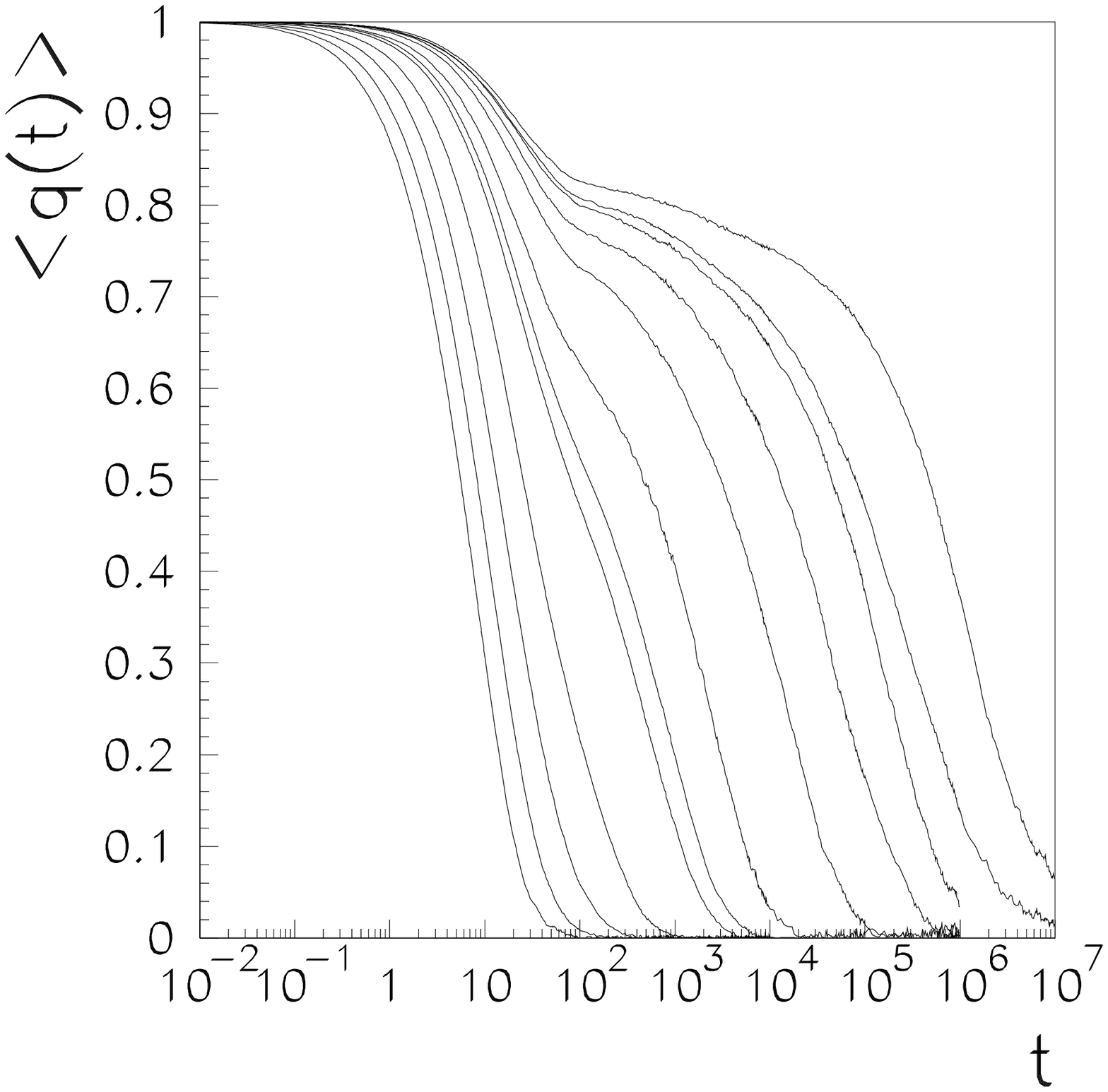}}
\end{picture}}
\ec
\caption{a) Mean square displacement $\lan r^2(r)\ran$ and b)
self overlap $\lan q(t)\ran$ as a function of time, in a three-dimensional
system of size $15^3$, for densities $\rho=0.55$, 0.6, 0.65, 0.7, 0.75, 0.76,
0.78, 0.8, 0.81, 0.815, 0.818, 0.821.}
\label{relax}
\efig

\bfig
\bc
\mbox{\unitlength=1cm
\begin{picture}(8,4)(0.3,0)
\put(0.4,0.2){\bf a}
\put(0,0){\epsfysize=4.5cm\epsfbox{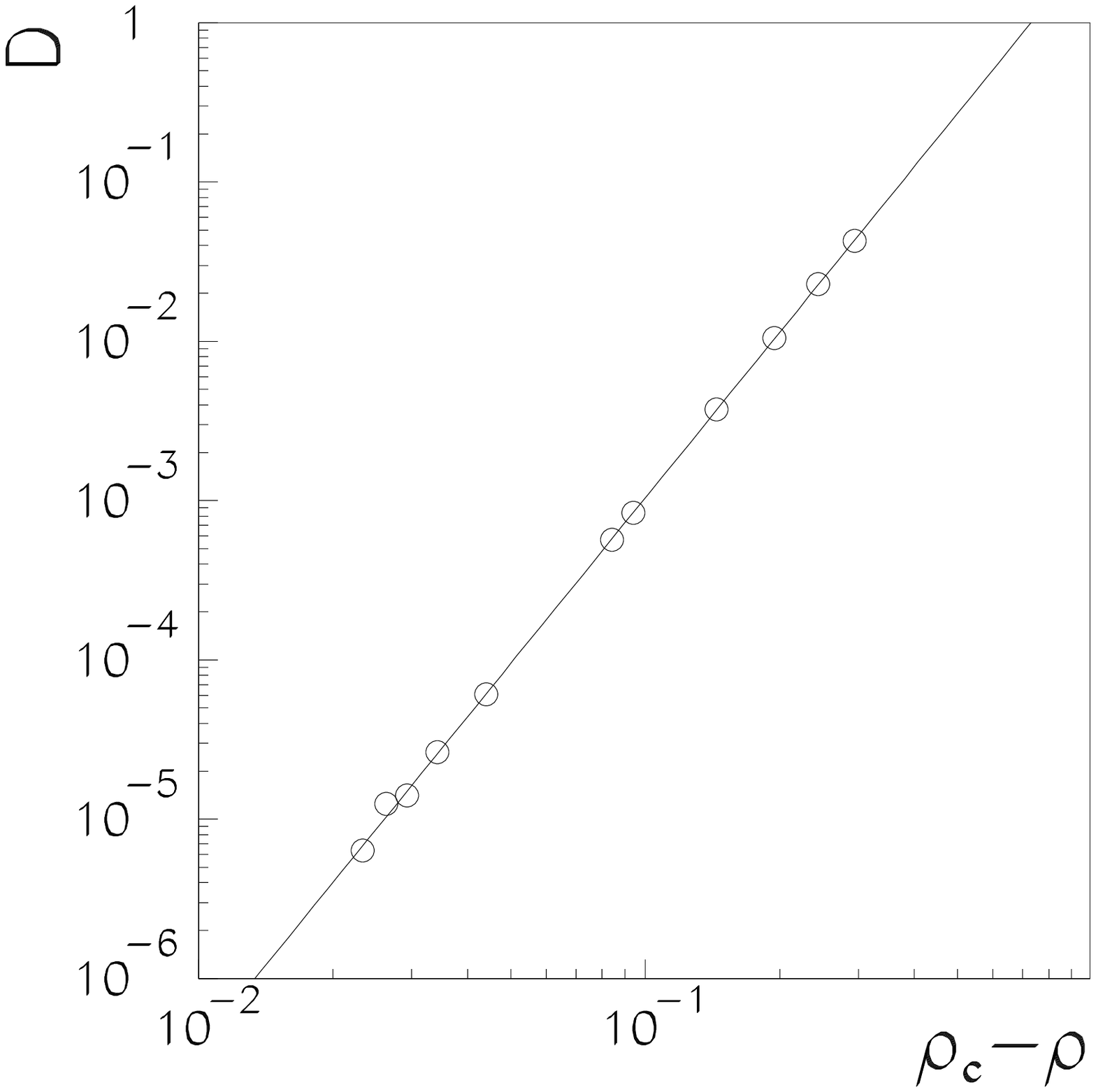}}
\put(4.8,0.2){\bf b}
\put(4.4,0){\epsfysize=4.5cm\epsfbox{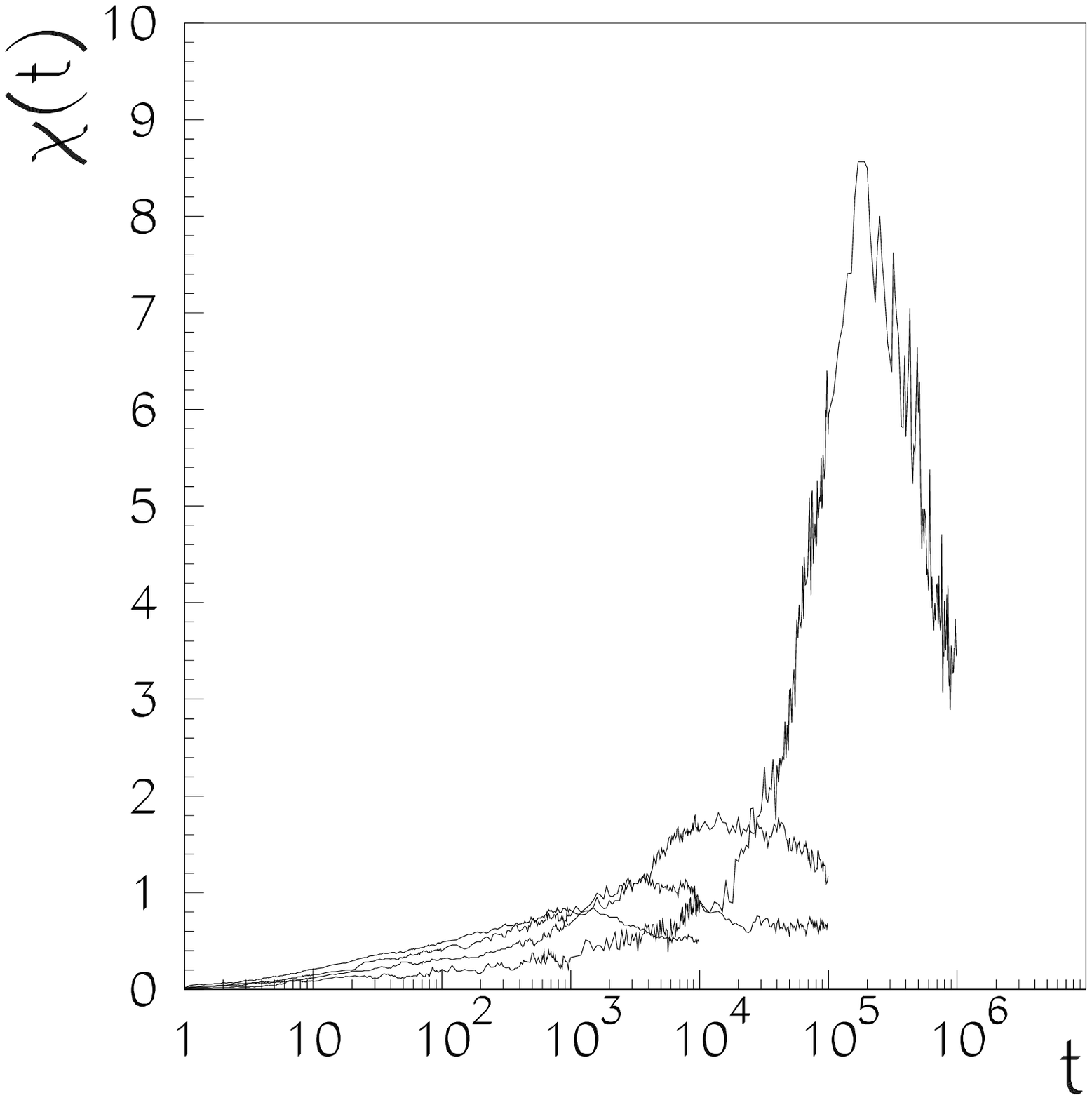}}
\end{picture}}
\ec
\caption{a) Diffusivity $D$ as a function of $\rho_c{-}\rho$,
with critical density $\rho_c=0.844$. The fitting function is a power law
$D=a|\rho-\rho_c|^\gamma$, with $\gamma=3.45$.
b) Dynamical susceptibility $\chi(t)$ as a function of time,
for densities $\rho=0.76$, 0.78, 0.8, 0.818.}
\label{plaw}
\efig

\bfig
\bc
\mbox{\unitlength=1cm
\begin{picture}(8,4)(0.3,0)
\put(0.4,0.2){\bf a}
\put(0.4,0.4){\epsfysize=4cm\epsfbox{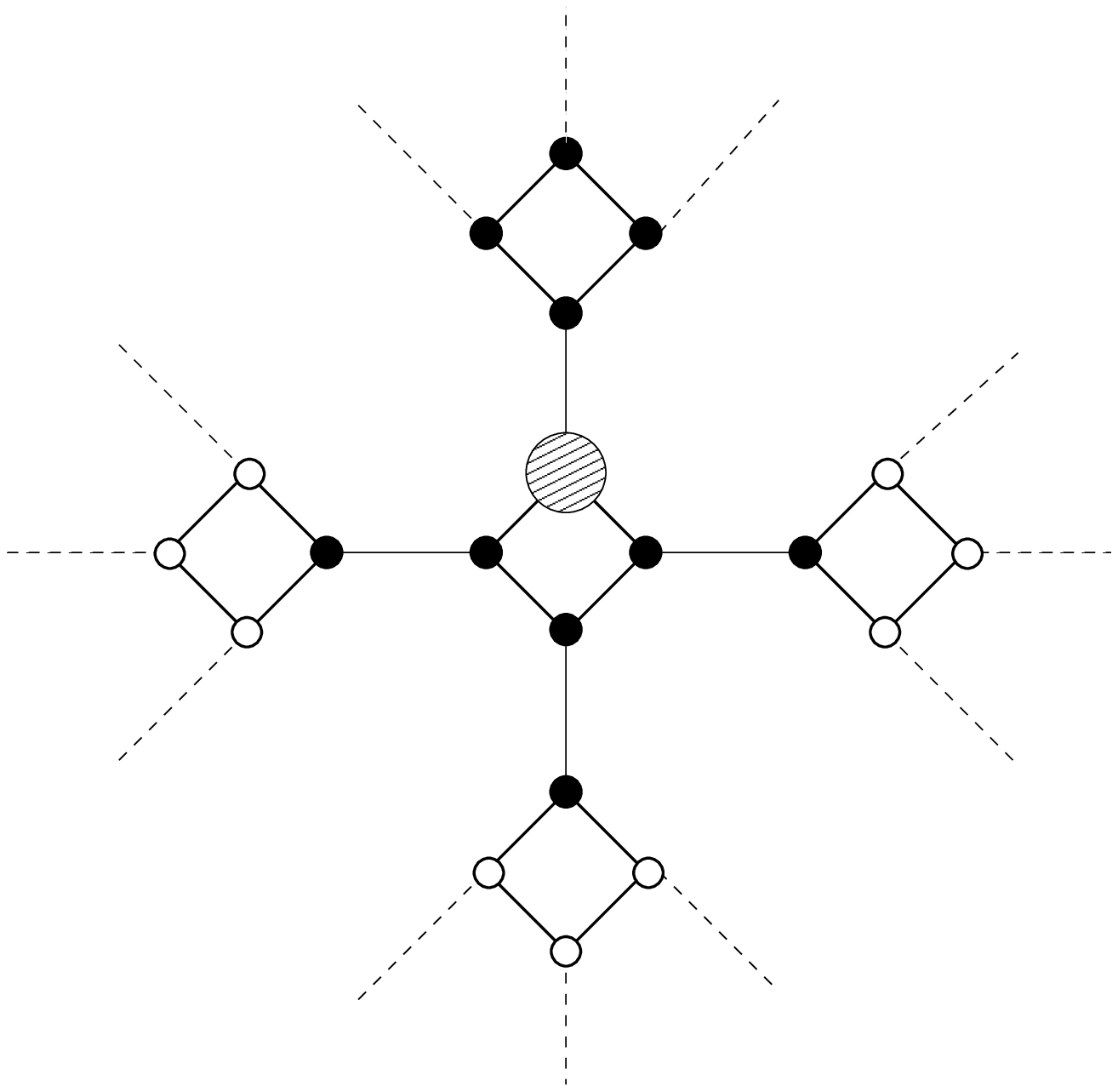}}
\put(4.8,0.2){\bf b}
\put(4.8,0.4){\epsfysize=3.5cm\epsfbox{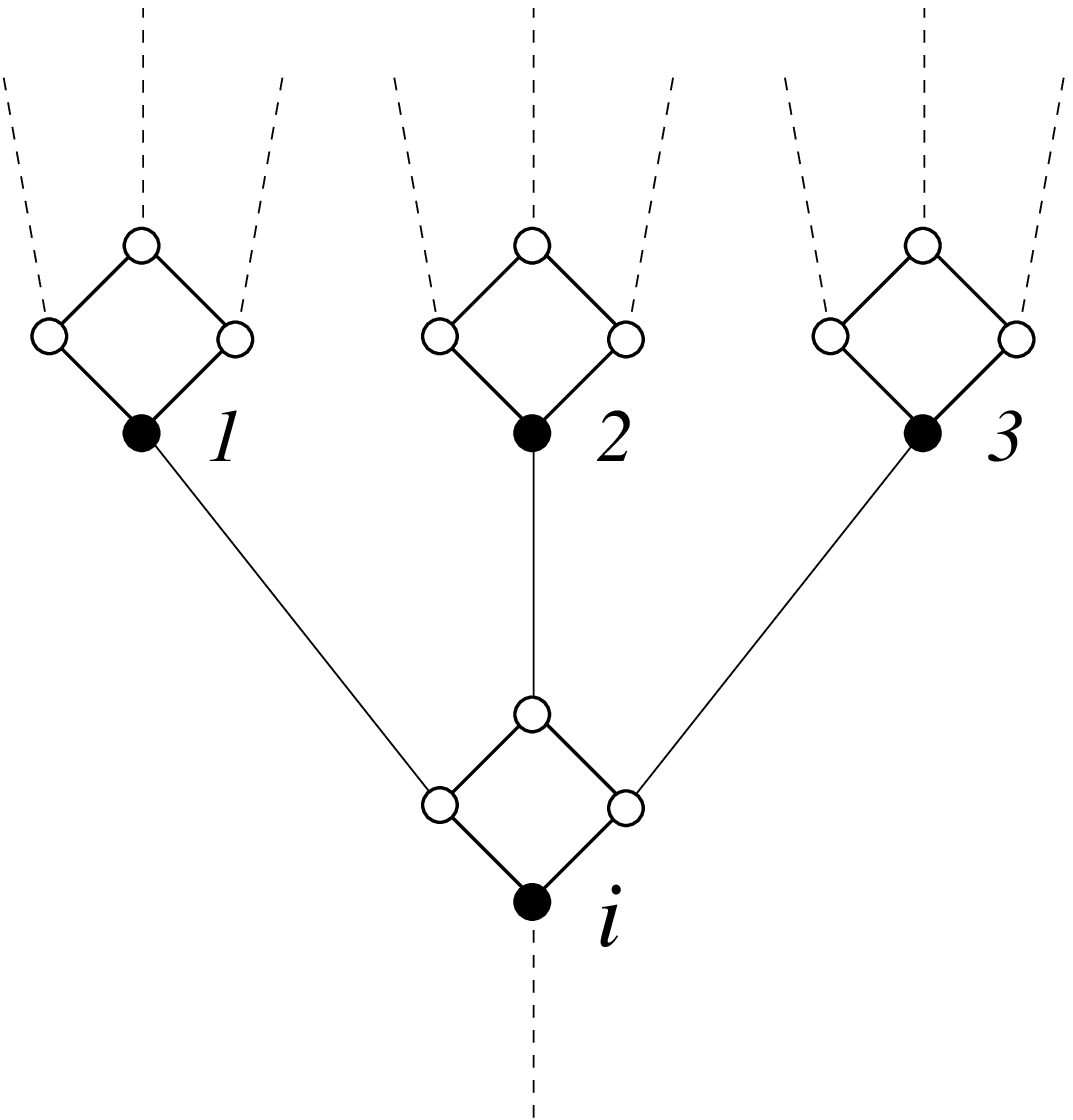}}
\end{picture}}
\ec
\caption{a) The model on the Bethe lattice (here with $k=3$):
each site is subdivided in $k+1$ positions (little circles),
and connected to $k+1$
randomly chosen neighbors.
A particle in a given position
(big shaded circle) forbids the presence of another particle in the
positions colored in black. b) Merging of the branches $j=1,\ldots,k$
onto the site $i$:
the ``external'' positions are those  colored in black,
the others are the ``internal'' ones.}
\label{tree}
\efig

\bfig
\bc
\mbox{\epsfysize=5.5cm\epsfbox{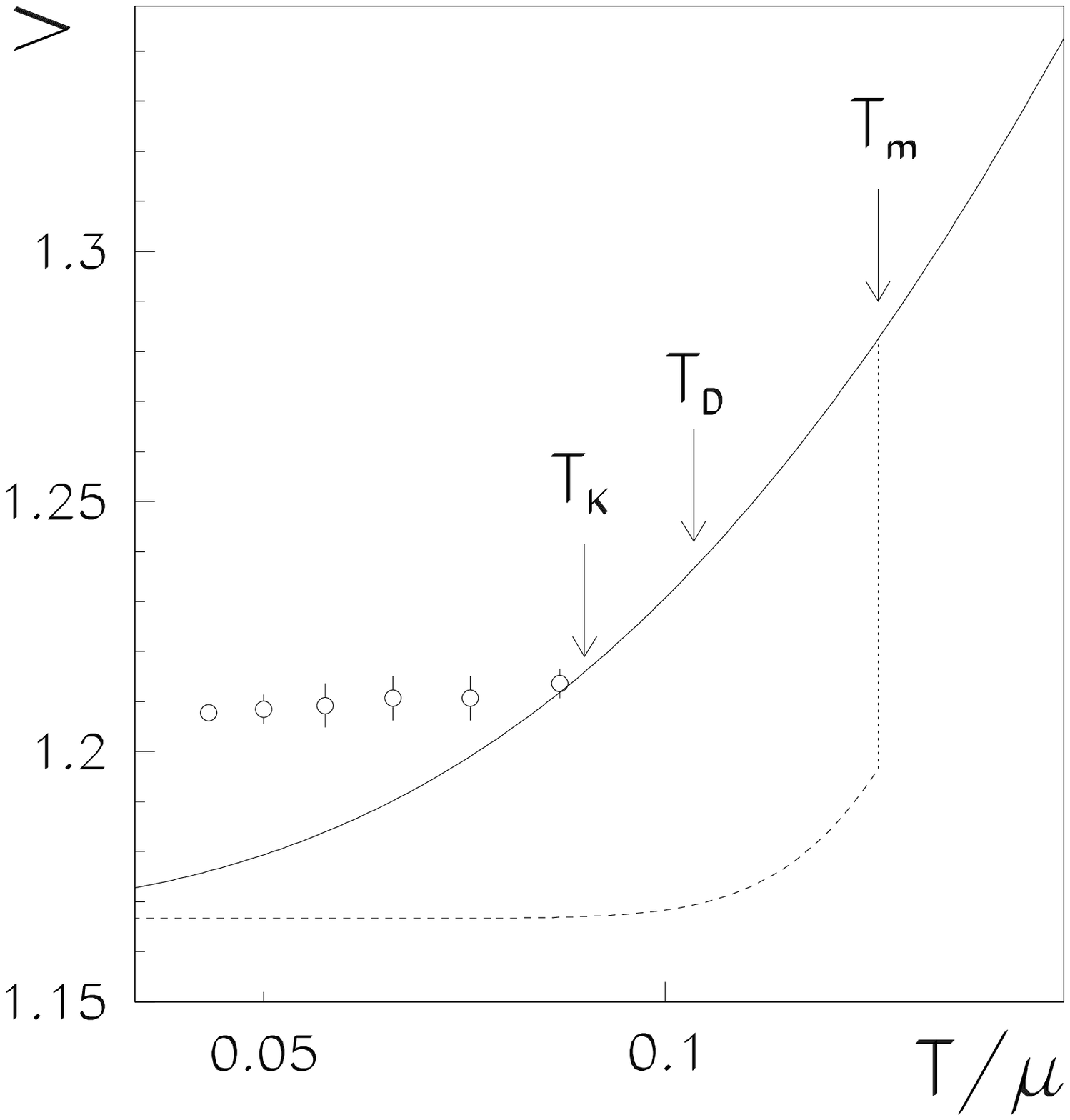}}
\ec
\caption{Specific volume (inverse density) as a function of
the temperature, on the Bethe lattice
with $k=5$. Solid line: replica symmetric liquid phase;
broken line: crystalline phase;
open circles: glassy phase in the 1-step RSB approximation.
Arrows mark the various temperatures cited in the text.}
\label{volume}
\efig

\bfig
\bc
\mbox{\epsfysize=5.5cm\epsfbox{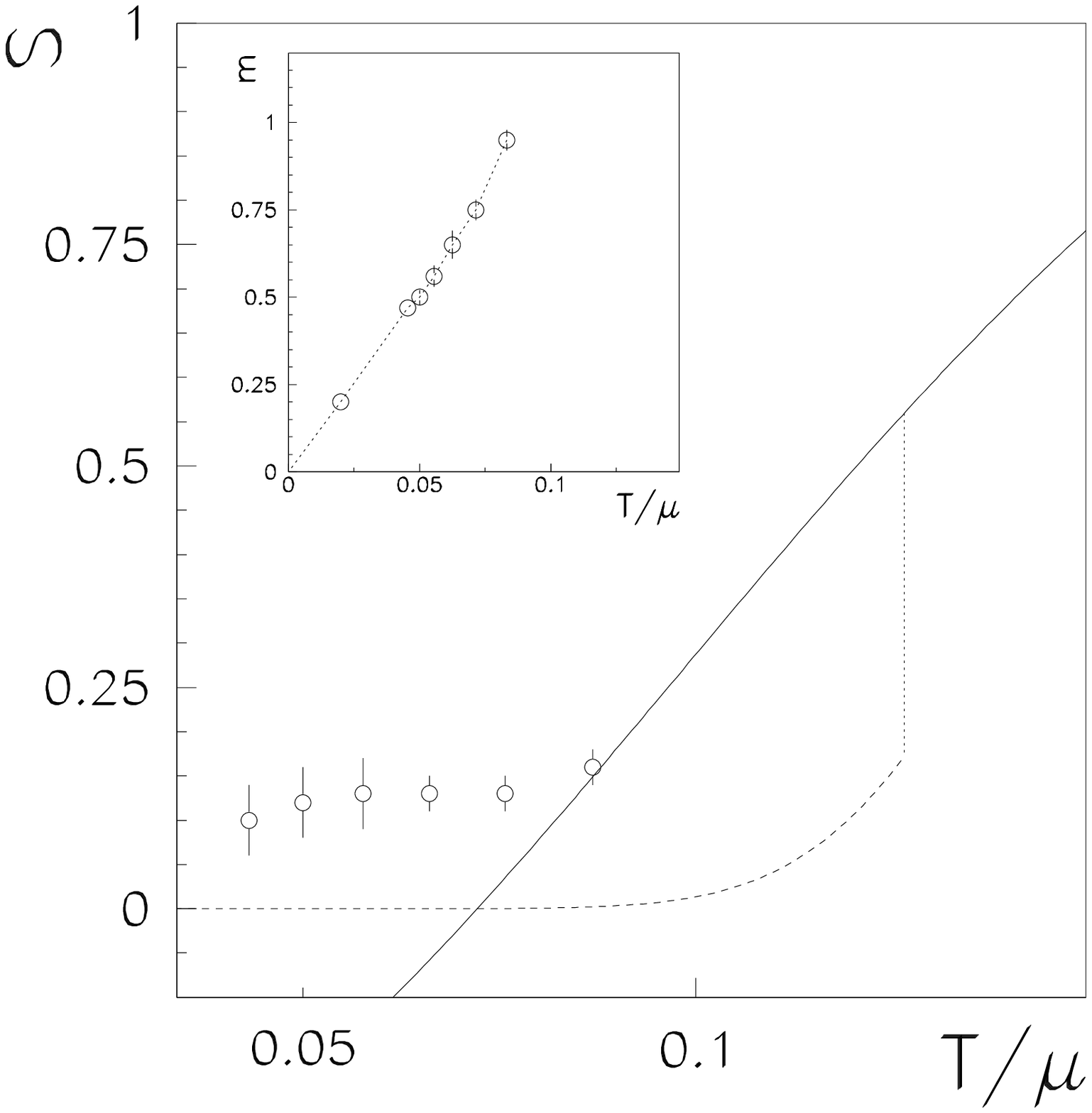}}
\ec
\caption{The same as in Fig.\ \protect\ref{volume},
but for the entropy per site.
Inset: parameter $m$ as a function of the temperature. The line is 
a guide for the eye.}
\label{entropy}
\efig

\end{document}